\documentclass[12pt,twoside]{article}
\usepackage{xrb2007}
\usepackage{psfig}
\usepackage{epsf}

\begin{document}

\session{Faint XRBs and Galactic LMXBs}

\shortauthor{Cornelisse et~al.}
\shorttitle{The Bowen Survey}

\title{An Overview of the Bowen Survey; detecting donor star signatures in Low Mass X-ray Binaries}
\author{Remon Cornelisse, Jorge Casares, Teo Mu\~noz-Darias}
\affil{Instituto de Astrofisica de Canarias, Via Lactea, La Laguna E-38200, Santa Cruz de Tenerife, Spain}

\author{Danny Steeghs}
\affil{Department of Physics, University of Warwick, Coventry, CV4 7AL, UK}
\author{Phil Charles}
\affil{South Africa Astronomical Observatory, P.O.Box 9.Observatory 7935, South
Africa\\
School of Physics and Astronomy, University of Southampton, Highfield, Southampton SO17 1BJ, UK}
\author{Rob Hynes}
\affil{Department of Physics and Astronomy, 202 Nicholson Hall, Louisiana State University, Baton Rouge, LA 70803, USA}
\author{Kieran O'Brien}
\affil{European Southern Observatory, Casilla 19001, Santiago 19, Chile}
\author{Andrew Barnes}
\affil{School of Physics and Astronomy, University of Southampton, Highfield, Southampton SO17 1BJ, UK}

\begin{abstract}
  In this paper we give a review of the Bowen fluorescence survey,
  showing that narrow emission lines (mainly N\,III and C\,III lines
  between 4630 and 4660 \AA) appear to be universally present in the
  Bowen blend of optically bright low mass X-ray binaries. These
  narrow lines are attributed to reprocessing in the companion star
  giving the first estimates of $K_2$, and thereby providing the first
  constraints on their system parameters.  We will give an overview of
  the constraints on the masses of the compact objects and briefly
  highlight the most important results of the survey.  Furthermore, we
  will point out the most promising systems for future follow-up
  studies and indicate how we think their estimates of the component
  masses can be improved.

\end{abstract}

\section{Introduction}

One of the main aims of optical observations of low-mass X-ray
binaries (LMXBs) has always been to find a signature of the donor
star, and thereby constrain the masses of both components.
Unfortunately, the optical emission of most LMXBs that accrete at high
rates is completely dominated by the reprocessing of the X-rays in the
outer accretion disk. This is the reason why, in spite of optical
counterparts being known for persistent LMXBs for years (sometimes
even over 20 years), no strong constraints on their system parameters
existed until recently.

\begin{figure*}[t]
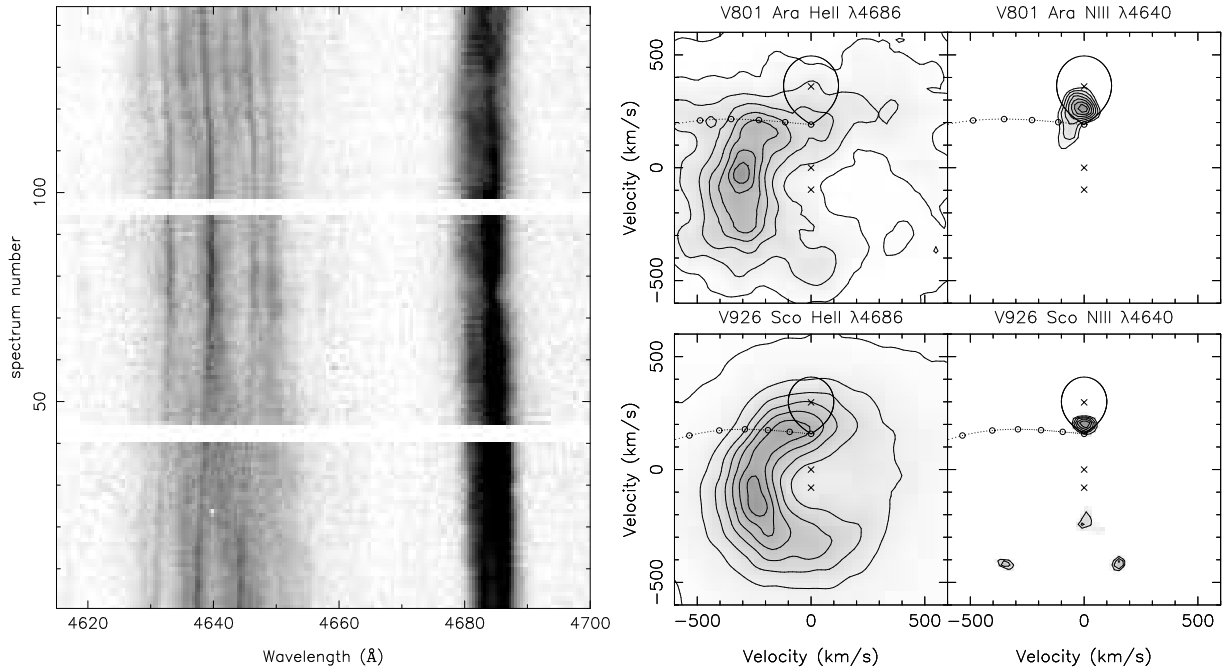

\parbox{8.0cm}{\psfig{figure=sco.ps,width=8.0cm}}
\parbox{8.0cm}{\psfig{figure=doppler.eps,angle=-90,width=8.0cm}}
\caption{{\it Left:} Trailed spectrogram of the Bowen blend and He\,II 
$\lambda$4686 of Sco\,X-1 showing the presence and the movement of the 
narrow emission lines. From \citet{sc02}. {\it Right:} Example Doppler maps
of  He\,II $\lambda$4686 (left) and the Bowen region (right) for 
4U\,1636$-$536 (top) and 4U\,1735$-$44 (bottom). The He\,II maps trace the
accretion disk, while the Bowen maps are dominated by a compact spot at the
phasing and velocity where the companion star is expected. The Roche lobe of 
the secondary and the gas stream trajectory are overplotted for clarity. From
\citet{ccs06}.
\label{scox1}}
\end{figure*}

The discovery of narrow high-excitation emission lines in the bright
LMXB Sco\,X-1 opened new opportunities for system parameter
constraints in active X-ray binaries \citep{sc02}.  Phase-resolved
blue spectroscopy showed that these narrow lines moved in anti-phase
with the compact object, strongly suggesting that they arise from the
irradiated surface of the donor star. This lead to the first estimate
of the semi-amplitude of the radial velocity of the secondary, $K_2$,
and mass function ($f$($M_1$)=$K_2^3$$P_{\rm orb}$/4$\pi$$G$) of
Sco\,X-1. These lines were most prominent in the Bowen region (the
region from 4630 to 4660 \AA), that mainly consists of a blend of
N\,III and C\,III lines. The N\,III lines are the result of a UV
fluorescence process, while the C\,III lines are due to
photo-ionization and subsequent recombination \citep{mct75}.  In this
paper we review the results of a survey we performed to apply this new
technique to other persistent LMXBs or transients during outburst in
order to find a donor star signature.

\section{The observed sources}

After the detection of the narrow emission lines in the Bowen region
in Sco\,X-1 (see Fig.\,\ref{scox1}/Left), a systematic survey was
performed targetting active LMXBs with known optical counterparts
brighter than $V$$\simeq$20. Thus far, phase resolved
spectroscopy has been performed for 7 persistent and 2 transient
LMXBs, using either the WHT, AAT, NTT or VLT telescope. In all cases a
signature of the donor star was detected, leading to a first estimate
of their mass functions. We have listed the observed sources plus the
mass constraints of their compact objects in Table\,\ref{limits}.

\begin{table}[t!]\begin{center}
\caption{Overview of all LMXBs that we have studied thus far. Indicated 
are the compact object mass limits, derived from
the narrow Bowen features, for each source. The transient sources
in our sample are indicated with an asterisk (*).
\label{limits}}
\begin{tabular}{lcl}
\hline
Source & Mass limits & reference\\
       & ($M_\odot$)\\ 
\hline
Sco\,X-1         & $\ge$0.05     & \citet{sc02}\\
GX\,339$-$4$^*$  & $\ge$5.8      & \citet{hsc03}\\
X1822$-$371      & $\ge$1.61     & \citet{csh03}\\
                 &               & \citet{mcm05}\\
4U\,1636$-$536   & $\ge$0.76$\pm$0.47 & \citet{ccs06}\\
4U\,1735$-$444   & $\ge$0.53$\pm$0.44 & \citet{ccs06}\\
Aql\,X-1$^*$     & $\ge$1.2      & \citet{ccs07a}\\
4U\,1254$-$69    & 1.20-2.64     & \citet{bcc07}\\
GX\,9$+$9        & $\ge$0.22     & \citet{csc07b}\\
LMC\,X-2         & $\ge$0.86     & \citet{csc07c}\\
\hline
\end{tabular}
\end{center}\end{table}

For most systems the narrow emission lines are too faint to be
detectable in the individual spectra. Fortunately, thanks to the
technique of Doppler tomography it is possible to use the information
contained in all phase resolved spectra at once to reconstruct the
emission distribution of the Bowen emission lines in velocity space.
In all cases this produced a compact spot consistent with the phasing
and the velocity of the companion star (for an example see
Fig.\,\ref{scox1}/Right). 

Apart from the problems that are inherent to all other methods to
estimate the system parameters of LMXBs (such as not knowing the
inclination or the radial velocity semi-amplitude of the primary,
$K_1$, for most sources), the Bowen technique has another
disadvantage. The velocity determined from the narrow components is
only a lower-limit to the true $K_2$ velocity, since these lines arise
on the irradiated side of the donor star which (for most of the
systems) does not correspond to the center of mass of the secondary
(note that this is also a problem for absorption lines if heating is
significant).  This displacement can be accounted for by the so-called
$K$-correction, which depends on the inclination of the system, the
opening-angle of the accretion disk, the mass ratio (see \citet{mcm05}).

\section{Highlights}

\begin{figure}[t]\begin{center}
\psfig{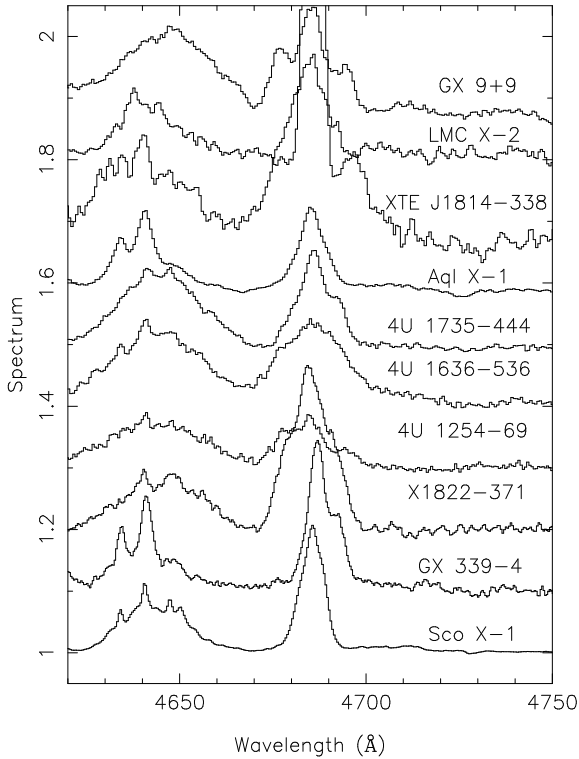}
\caption{Average Doppler corrected spectra of the Bowen region in the rest 
frame of the companion. Shown are the systems observed by our group thus far.
\label{spectra}}
\end{center}\end{figure}

Thus far, the survey has shown that all observed systems show narrow
emission lines in the Bowen region that can be attributed to the
irradiated surface of the companion. In Fig.\,\ref{spectra} we show
Bowen region of the average Doppler corrected spectra in the rest
frame of the companion, and in all spectra narrow N\,III and/or C\,III
lines are visible. This suggests that this technique can be
universally applied, and is an excellent tool to constrain the system
parameters of LMXBs.  Not only does it appear to work for persistent
sources, but GX\,339$-$4 and Aql\,X-1 have shown that it can also work
for transients during outburst. This technique can therefore also be
used to constrain the system parameters for transients that are too
faint in optical when they return to quiescence.

The observation of GX\,339$-$4 is a spectacular demonstration of the
power of the Bowen fluorescence technique. Despite being one of the
earliest proposed black hole candidates even its orbital period was
uncertain due to the bright accretion disk that still dominates the
optical flux when the system is in quiescence.  This made it not
possible to directly measure the system parameters of this system thus
far.  After applying the Bowen technique to GX\,339$-$4 a mass
function of 5.8$M_\odot$ was found, finally confirming the black hole
nature of its compact object \citep{hsc03}.

Another interesting case is the neutron star transient Aql\,X-1. In
quiescence the counterpart is 2 magnitudes fainter than a nearby star,
severely hampering quiescent studies. During a 2004 outburst we had
the opportunity to take several high resolution blue spectra, and the
narrow emission components were clearly present in the individual
spectra. The resulting radial velocity curve lead to a lower limit on
the mass function of $f(M_1)$$\ge$1.23$\pm$0.12$M_\odot$. Combining
this with estimates for the rotational broadening of these lines
suggests that the mass of the compact object is $\ge$1.6$M_\odot$ (at
95\%), strongly implying a massive neutron star in Aql\,X-1
\citep{ccs07a}

Not only does this technique work for Galactic LMXBs, but it has also
been successfully applied to the extra-galactic system LMC\,X-2
\citep{csc07c}. From the He\,II $\lambda$4686 emission line it was
possible to determine a spectroscopic period of 0.32$\pm$0.02 day,
that was interpreted as the orbital period. However, the spectra do
show a longer term variation that is most likely due to a precessing
accretion disk. Despite these difficulties, plus the fact that the
Bowen emission is much weaker compared to the Galactic systems, it was
possible to detect the narrow lines using Doppler tomography.

As already pointed out, a main problem in determining the system
parameters from dynamical studies of LMXBs is the lack of knowledge of
$K_1$ and the inclination. However, for several systems these
parameters are known, making them excellent system to derive strong
constraints on the component masses. For 4U\,1254$-$69 the inclination
is well constrained due to the presence of dips, while $K_1$ is known
for the neutron star LMXB 4U\,1636$-$53. However, the most promising
system is the eclipsing pulsar X1822$-$371, for which both the
inclination and $K_1$ are known. By accurately modelling the narrow
components for this system it should be possible to strongly constrain
both the donor and neutron star mass (Mu\~noz-Darias et~al.  2008 in
prep.).

 \section{Conclusions}

 We have given a short overview of the novel technique of Bowen
 fluorescence that has proven itself to work for the 9 systems thus
 far observed (see Table\,\ref{limits}), and for most of the systems
 we have been able to derive the first constraints on their system
 parameters ever.  There are still a handful of persistent LMXBS left
 that are bright enough for this technique, and there are plans to
 also observe these systems. Recently 4U\,1957$+$11 has been observed
 with Magellan, and in the near future EXO\,0748$-$676, Ser\,X-1 and
 4U\,1556$-$605 will be observed with the VLT. Furthermore, this
 technique might also be an excellent tool to determine the system
 parameters of future bright transients.

 The next step will be to improve the determined $K_2$ velocities for
 the most promising candidates by measuring the distribution of the
 emission across the Roche-lobe using high resolution spectroscopy.
 This will allow us to more accurately measure the radial velocities
 of the narrow lines, determine their rotational broadening, and
 hopefully model the shape of the lines as a function of orbital
 period in order to get a better constraint on the $K$-correction.
 Furthermore, we have started using the fact that the Bowen lines are
 efficiently reprocessed on the donor star to constrain the
 inclination and the $K$-correction using Echo-tomography with a
 narrow band filter centered around the Bowen blend (for more
 information see Mu\~noz-Darias et~al. in this volume). These next
 steps will hopefully provide stronger constraints on the masses of
 LMXBs, and in particular give accurate neutron star masses.

\end{document}